\documentstyle[12pt,epsfig]{article}

\font\ten=cmbx10 at 12pt

\renewcommand{\thefootnote}{\fnsymbol{footnote}}

\newcounter{saveeqn}

\newenvironment{goodlist}[1]%
{\begin{list}{}{\settowidth{\labelwidth}{#1}
  \setlength{\leftmargin}{\labelwidth}
  \addtolength{\leftmargin}{\labelsep}
  \setlength{\parsep}{1ex plus0.7ex minus0.7ex}
  \setlength{\itemsep}{0.8ex}
  }}{\end{list}}

\def\build#1_#2^#3{\mathrel{\mathop{\kern 0pt#1}\limits_{#2}^{#3}}}

\begin{document}

\begin{titlepage}

\begin{center}

{\ten Centre de Physique Th\'eorique\footnote{Unit\'e Propre de
Recherche 7061} - CNRS - Luminy, Case 907}

{\ten F-13288 Marseille Cedex 9 - France }

\vspace{1 cm}

{\large\bf FURTHER EXPERIMENTAL TESTS FOR SIMPLE RELATIONS
BETWEEN UNPOLARIZED AND POLARIZED QUARK PARTON DISTRIBUTIONS}

\vspace{0.3 cm}
\setcounter{footnote}{0}
\renewcommand{\thefootnote}{\arabic{footnote}}

{\bf Claude BOURRELY and Jacques SOFFER}

\vspace{1 cm}

{\bf Abstract}

\end{center}

Some simple relations between unpolarized and polarized quark parton
distributions have direct experimental consequences which will be
presented here. In particular, we will see that it is possible to
relate the deep inelastic structure functions $F_2$ and $g_1$, both
for proton and deuteron, in fair agreement with experimental data.

\vspace{1 cm}

\noindent Key-Words : polarized parton densities, deep-inelastic
scattering.

\bigskip

\noindent Number of figures : 3

\bigskip

\noindent August 1995

\noindent CPT-95/P.3224

\bigskip

\noindent anonymous ftp or gopher: cpt.univ-mrs.fr

\end{titlepage}

Deep inelastic scattering of polarized electrons and muons off
polarized targets have been used to study the internal structure of the
nucleon. The most abundant and accurate experimental information we
have so far, concerns the so called $g_1$ spin structure function
obtained with longitudinally polarized leptons on longitudinally
polarized protons, deuterons and $^3He$ targets
\cite{SMCa,E143a,E143b,SMCb,E142}
which allow to get separately $g^p_1$ for protons and $g^n_1$ for
neutrons. In the scaling limit, $g_1$ is a function of only the well
known Bjorken variable $x = Q^2/2m_p \nu$, but when $Q^2$, the square
of the four-momentum of the virtual photon of energy $\nu$, is finite,
$g_1$ depends also on $Q^2$. So let us consider a situation with
scaling violations and we recall that for protons one has
\begin{equation}\label{g1p}
g^p_1(x, Q^2) = \frac{1}{2}\left[ \frac{4}{9}\Delta u(x, Q^2) +
\frac{1}{9} \Delta d(x, Q^2) + { 1 \over 9} \Delta s(x, Q^2) \right],
\end{equation}
where for each flavor the polarized quark distribution
\begin{equation}\label{delq}
\Delta q(x, Q^2) = q_+(x, Q^2) - q_-(x, Q^2),
\end{equation}
includes both quark and antiquark contributions. One obtains the
neutron structure function $g^n_1(x, Q^2)$ by exchanging $u$ and $d$
and for the deuteron one has
\begin{equation}\label{g1d}
g^d_1 = { 1 \over 2}(g^p_1 + g^n_1)(1 - 3/2 \omega_D)
\end{equation}
where $\omega_D = 0.05 \pm 0.01$ is the probability of the deuteron
to be in a $D$-state. The measurements of polarized deep inelastic
scattering yield in fact the spin asymmetry $A^N_1$ which is, to a
good approximation,
\begin{equation}\label{a1n}
A^N_1(x, Q^2) = {2 x g^N_1(x, Q^2) \over F^N_2(x, Q^2)/
(1 + R^N(x, Q^2))},
\end{equation}
where $N = p,n$ or $d$, $F^N_2(x, Q^2)$ is the
corresponding unpolarized structure function and
$R^N = (F^N_2 - 2 x F^N_1)/2 x F^N_1$ is small or zero in the leading
order approximation. It is a well known experimental fact that
$A^N_1(x, Q^2)$ is $Q^2$ independent, in the kinematical range so far
 explored. This means that numerator and denominator in Eq. (\ref{a1n})
have the same $Q^2$ dependence, within the present experimental
accuracy. Clearly this observation, which has no sound theoretical
justification, can be however used for phenomenological purposes,
as we will see in what follows.

Sometime ago \cite{BUS,BS1,BS2} we have advocated the exclusion
Pauli principle to postulate the existence of simple relations
between unpolarized and polarized quark parton distributions.
In particular, we have assumed \cite{BUS,BS1,BS2} that
\begin{equation}\label{delu}
\Delta u(x, Q^2) = u(x, Q^2) - d(x, Q^2)
\end{equation}
and also that the $s$ quarks are unpolarized ({\it i.e.} $\Delta s(x, Q^2)
 = 0$). In addition, for the moment, if we ignore the contribution of
$\Delta d(x, Q^2)$ in $g^p_1$, we obtain the simple relationship
\begin{equation}\label{xg1p}
xg^p_1(x, Q^2) = \frac{2}{3} \left[ F^p_2(x, Q^2) - F^n_2(x, Q^2)
 \right] .
\end{equation}
As a result for the proton case Eq. (\ref{a1n}) reads
\begin{equation}\label{a1p}
A^p_1(x, Q^2) = {4 \over 3}( 1 - r(x, Q^2))(1 + R^p(x, Q^2))
\end{equation}
where $r(x, Q^2) = F^n_2(x, Q^2)/F^p_2(x, Q^2)$.

The NMC-NA37 experiment at CERN \cite{NMCa} has performed a very
detailed measurement of $F^p_2$ and $F^d_2$ in a large kinematic
range $0.006 \leq x \leq 0.6$ and $0.5 \leq Q^2 \leq 55GeV^2$ and
we have used their very new parametrization of the data \cite{BRU}
to calculate $r(x, Q^2)$. The best determination of $R^p(x, Q^2)$ in
the kinematic range $0.1 \leq x \leq 0.9$ and $0.6 \leq Q^2 \leq
20GeV^2$ can be found in \cite{WHIT}. Therefore one can make a
numerical estimate of Eq. (\ref{a1p}) which gives the contribution
of $u$ quarks only to $A^p_1(x, Q^2)$. The results are shown in
Fig.1 for two $Q^2$ values, $Q^2 = 3$ and $10GeV^2$, together with
the proton data from SMC\cite{SMCa} and SLAC\cite{E143a}. It is
remarkable to note that Eq. (\ref{a1p}) leads to almost no $Q^2$
dependence, except for $x < 0.1$ where it might be due to the lack
of a precise knowledge of $R^p$\cite{CCFR}. We also see that
Eq. (\ref{a1p}) gives the correct shape of the data, although it lies
systematically above it. This is easily understood because we have
neglected the contribution of the $d$ and $s$ quarks which are
expected to be negative. Actually one could try to estimate it from
the shift one observes with respect to the data, but one would not
be able to disentangle the $d$ quarks from the $s$ quarks. So we
propose instead, to use a different method which involves both proton
and deuteron data. As we have seen above the $g_1$ structure functions
for proton, neutron and deuteron are expressed in terms of the three
polarized distributions $\Delta u$, $\Delta d$ and $\Delta s$.
Clearly one can eliminate $\Delta s$ by considering $g^p_1 - g^n_1$
and similarly, one can eliminate $\Delta d$ by considering
$4 g^p_1 - g^n_1$. If we now assume $\Delta s(x, Q^2) = 0$ and
using Eq. (\ref{a1n}) we obtain as a consequence of Eq. (\ref{delu})
\begin{equation}\label{relat1}
5xg^p_1(x, Q^2) - \frac{4}{2 - 3\omega_D}xg^d_1(x, Q^2) =
 5 \left[ F^p_2(x, Q^2) - F^d_2(x, Q^2) \right] .
\end{equation}
This simple relationship between $g^p_1$, $g^d_1$ and $F^p_2$,
$F^d_2$ can be tested directly from experimental data. This test
is shown in Fig.2 where we have used for the {\it l.h.s.} of
Eq. (\ref{relat1}) the SMC data\cite{SMCa,SMCb} and the SLAC
data\cite{E143a,E143b} on the $g_1$'s and for the {\it r.h.s.}
the NMC parametrization\cite{BRU} for $F^p_2$ and $F^d_2$.
The errors on the {\it l.h.s.} have been calculated by using the
statistical errors only, added in quadrature. The estimated errors
for the {\it r.h.s.} are represented by dashed lines on both sides
of the full line which corresponds to the NMC parametrization.
The test is indeed very well satisfied and gives, within the
experimental uncertainty, a fairly good support to Eq. (\ref{delu})
and to the assumption $\Delta s(x, Q^2) = 0$. Moreover if one takes
the first moment of both sides of Eq. (\ref{relat1}), one finds for
the {\it l.h.s.} $0.588 \pm 0.054$ using the SLAC data or
$0.605 \pm 0.074$ using the SMC data, whereas the {\it r.h.s.}
gives $0.587 \pm 0.065$ using \cite{NMCb}.

Finally, let us consider the combination $4g^n_1 - g^p_1$ which
eliminates $\Delta u$. Again if we assume $\Delta s(x, Q^2) =0$,
by using Eq. (\ref{g1d}) one finds
\begin{equation}\label{relat2}
\frac{16}{2 - 3\omega_D}xg^d_1(x, Q^2) - 5xg^p_1(x, Q^2) =
 {5 \over 6}x\Delta d(x, Q^2).
\end{equation}
Now according to the arguments we have used in \cite{BS2}, one
is led to assume that
\begin{equation}\label{deld}
\Delta d(x, Q^2) = -{1 \over 3}d_{val}(x, Q^2).
\end{equation}
By combining Eqs. (\ref{relat2}) and (\ref{deld}) one gets the
following expression for $xd_{val}(x, Q^2)$
\begin{equation}\label{dval}
xd_{val}(x, Q^2) = \frac{18}{5} \left[ 5xg^p_1(x, q^2) -
\frac{16}{2 - 3\omega_D}xg^d_1(x, Q^2) \right]
\end{equation}
which also can be tested directly from the data on the $g_1$'s.
This test is shown in Fig.3 where, as previously, we have used
for the {\it r.h.s.} of Eq. (\ref{dval}) the SMC data\cite{SMCa,SMCb}
and the SLAC data\cite{E143a,E143b} on the $g_1$'s and for the
{\it l.h.s.} the following simple expression taken from \cite{BS2}
at $Q^2 = 3 GeV^2$
\begin{equation}\label{dvalb}
xd_{val}(x) = \frac{1.802 x^{0.738}}{e^{(x - 0.231)/0.092} + 1}.
\end{equation}
The experimental uncertainty in this case is indeed rather large
and, although this test does not lead to any inconsistency, we cannot
claim that it gives a very strong support to Eq. (\ref{deld}).

To conclude we would like to stress that our simple relations
between unpolarized and polarized quark parton distributions,
Eqs. (\ref{delu}) and (\ref{deld}), are consistent with present
experimental data. To establish them more firmly,
one needs a substancial improvement of the data which is
expected in the near future. Then it will be possible to decide
if these relations are just a useful phenomenological guide or if
they have a more fundamental significance.

\section*{Acknowledgments}
We thank G. Altarelli and R. Windmolders for interesting comments.
One of us (J.S.) is grateful for the kind hospitality at the Theory
Division (CERN) where part of this work was done.

\section*{Figure Captions}

\begin{goodlist}{Fig.3}

\item[Fig.1] The spin asymmetry $A^p_1(x, Q^2)$ versus $x$ calculated
from Eq. (\ref{a1p}) at $Q^2 = 3 GeV^2$ (dashed line) and $Q^2 = 10 GeV^2$
(solid line) compared to the data at different $Q^2$ (open squares
from \cite{SMCa} and full circles from \cite{E143a}).

\item[Fig.2] Experimental test of Eq. (\ref{relat1}). Open squares are
obtained from \cite{SMCa} and \cite{SMCb}, full circles from
\cite{E143a} and \cite{E143b}. Full line has been calculated
from the parametrization of Ref.\cite{BRU} at $Q^2 = 6GeV^2$ and
dashed lines are the corresponding estimated errors.

\item[Fig.3] Experimental test of Eq. (\ref{dval}). Open squares are
obtained from \cite{SMCa} and \cite{SMCb}, full circles from
\cite{E143a} and \cite{E143b}. Full line has been calculated
using Eq. (\ref{dvalb}) from \cite{BS2}.

\end{goodlist}

\newpage
\pagestyle{empty}
\begin{figure}[p]
\centerline{\epsfig{file=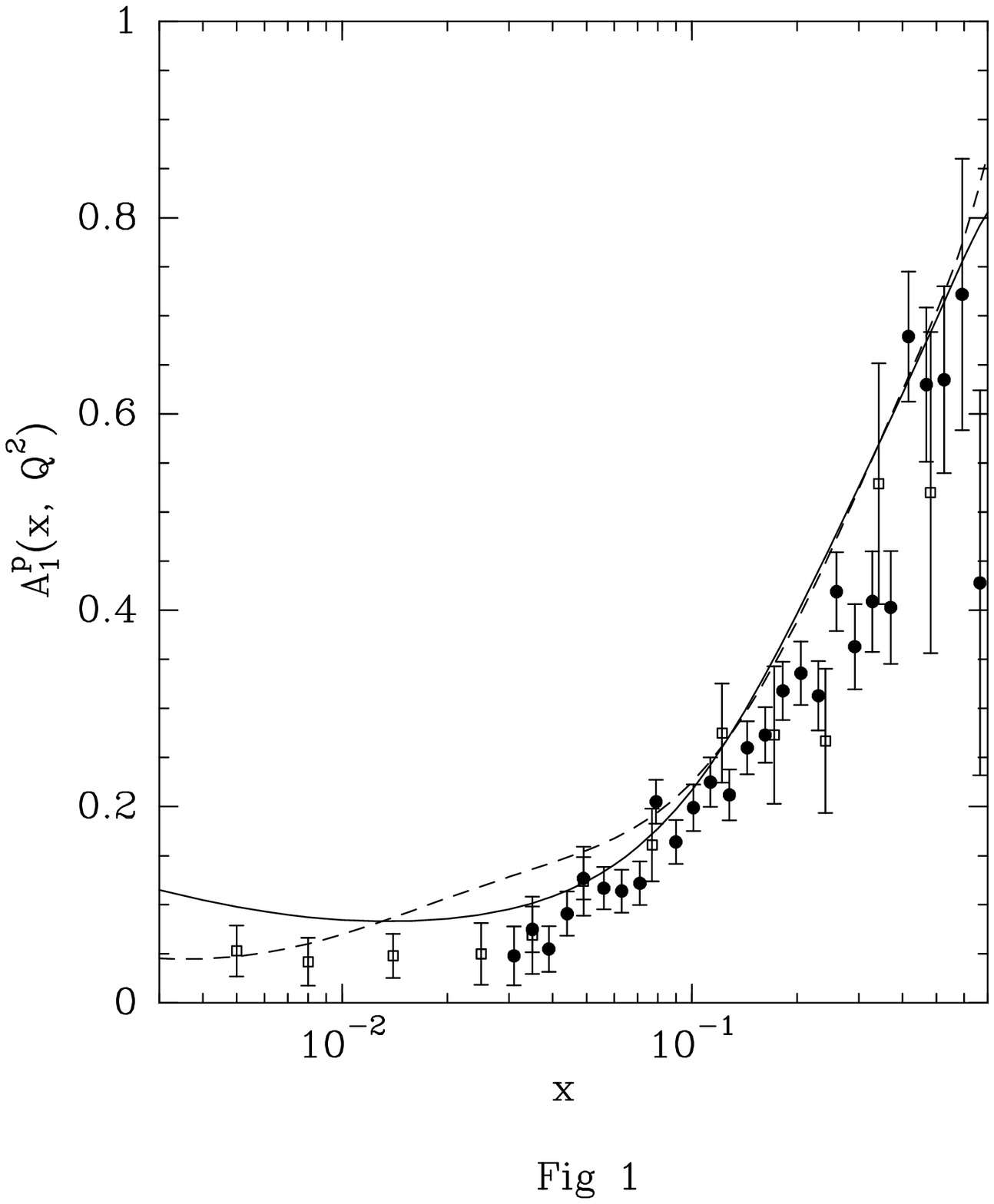}}
\label{fig:fig1}
\end{figure}

\newpage
\begin{figure}[p]
\centerline{\epsfig{file=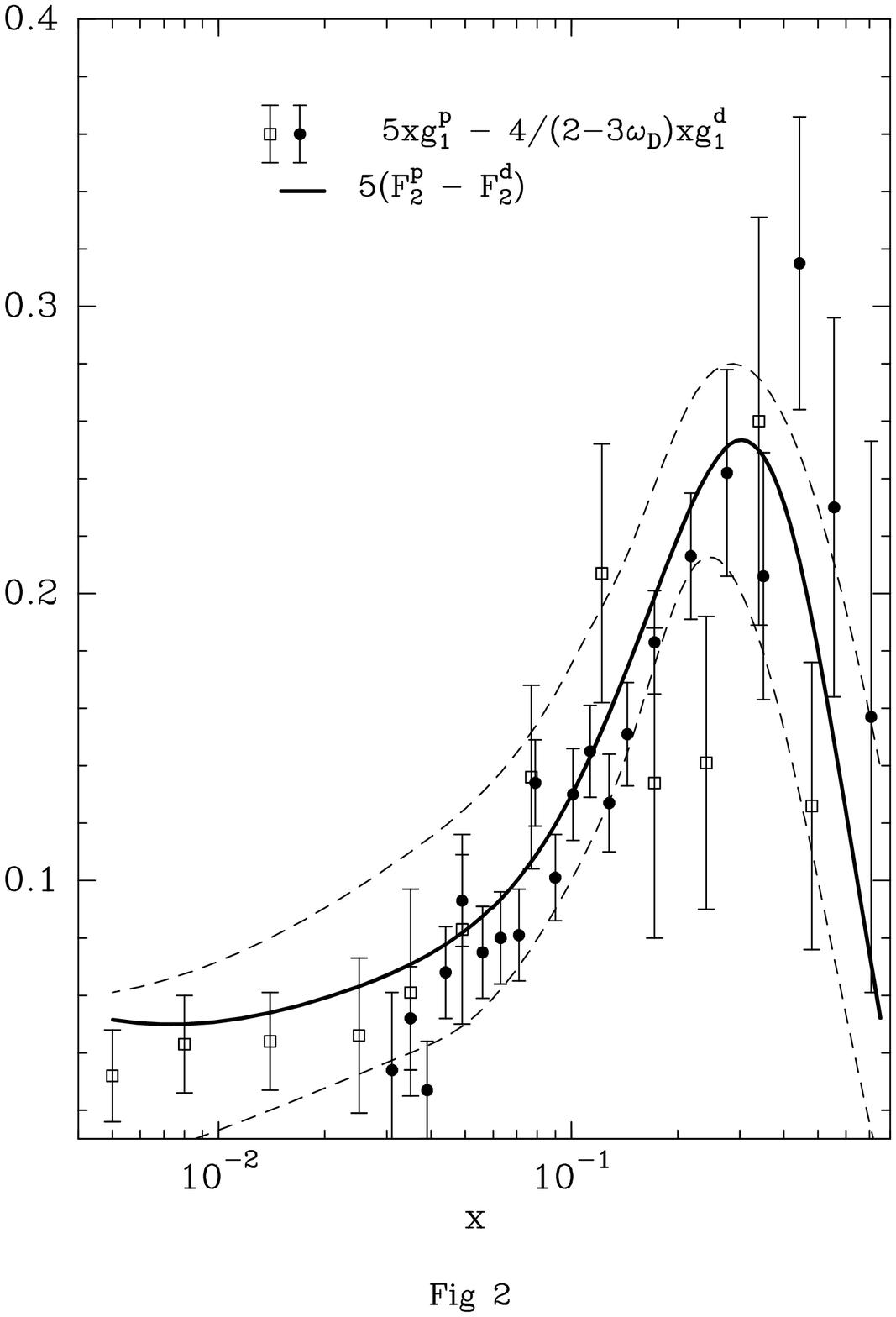}}
\label{fig:fig2}
\end{figure}

\newpage
\begin{figure}[p]
\centerline{\epsfig{file=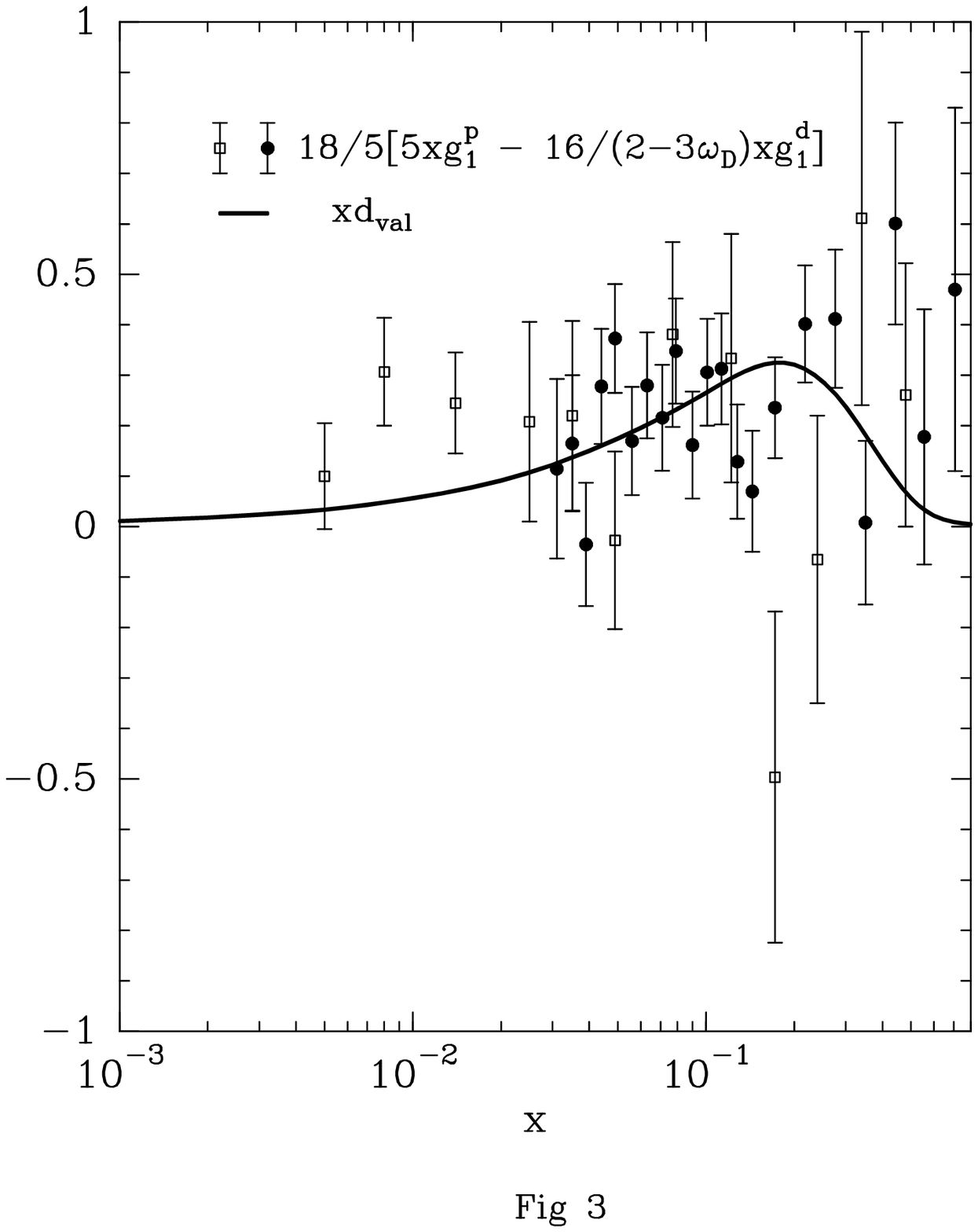}}
\label{fig:fig3}
\end{figure}


\begin{thebibliography}{99}

\bibitem{SMCa}Spin Muon Collaboration, D. Adams {\it et al.}, Phys.
Lett. {\bf B329}, 339 (1994); Erratum, Phys. Lett. {\bf B339},
332 (1994).

\bibitem{E143a}SLAC E143, K. Abe {\it et al.}, Phys. Rev. Lett.
{\bf 74}, 346 (1995).

\bibitem{E143b}SLAC E143, K. Abe {\it et al.}, preprint
SLAC-PUB95-6734, to appear in Phys. Rev. Lett..

\bibitem{SMCb}Spin Muon Collaboration, D. Adams {\it et al.},
Phys. Lett. {\bf B357}, 248 (1995).

\bibitem{E142} SLAC E142, D.L. Anthony {\it et al.}, Phys. Lett.
{\bf B259}, 959 (1993).

\bibitem{BUS}F. Buccella and J. Soffer, Mod. Phys. Lett., {\bf A8},
225 (1993); Europhysics Lett. {\bf 24}, 165 (1993); Phys. Rev.,
{\bf D48}, 5416 (1993).

\bibitem{BS1}C. Bourrely and J. Soffer, Phys. Rev. {\bf D51}, 2108
(1995).

\bibitem{BS2}C. Bourrely and J. Soffer, Nucl. Phys. {\bf B445},
341 (1995).

\bibitem{NMCa}New Muon Collaboration, P. Amandruz {\it et al.},
Phys. Lett. {\bf B295}, 159 (1992).

\bibitem{BRU}We thank A. Bruell for providing us with the new NMC
parametrization still preliminary. New Muon Collaboration, M.
Arneodo {\it et al.}, "Measurement of the proton and deuteron
structure functions $F^p_2$ and $F^d_2$", to be submitted to Phys. Lett.
{\bf B}.

\bibitem{WHIT}L.W. Whitlow {\it et al.}, Phys. Lett. {\bf B250},
193 (1990).

\bibitem{CCFR}A better parametrization of $R^p$ using CCFR neutrino
data at small values of $x$ will be available soon, (A. Bodek, private
communication).

\bibitem{NMCb} New Muon Collaboration, M. Arneodo {\it et al.},
Phys. Rev. {\bf D50}, R$1$ (1994).

\end{thebibliography}
\end{document}